\documentclass[aps,prb,twocolumn,superscriptaddress, show pacs]{revtex4-1}
\usepackage{bm, amsmath}
\usepackage{graphicx}
\usepackage{color}
\usepackage{epstopdf}
\usepackage{graphicx}
\usepackage{dcolumn}
\usepackage{bm}
\usepackage{subfigure}

\begin{document}

\title{Homogeneous and heterogeneous  nucleation of skyrmions in thin layers of cubic helimagnets}


\author{A. O. Leonov}
\thanks{leonov@hiroshima-u.ac.jp}
\affiliation{Chirality Research Center, Hiroshima University, Higashi-Hiroshima, Hiroshima 739-8526, Japan}
\affiliation{Department of Chemistry, Faculty of Science, Hiroshima University Kagamiyama, Higashi Hiroshima, Hiroshima 739-8526, Japan}
\affiliation{IFW Dresden, Postfach 270016, D-01171 Dresden, Germany} 


\author{K. Inoue}
\affiliation{Chirality Research Center, Hiroshima University, Higashi-Hiroshima, 
Hiroshima 739-8526, Japan}
\affiliation{Department of Chemistry, Faculty of Science, Hiroshima University Kagamiyama, Higashi Hiroshima, Hiroshima 739-8526, Japan}

\date{\today}

\begin{abstract}
{Formation of isolated chiral skyrmions by homogeneous and heterogeneous nucleation has been studied in thin layers of cubic helimagnets via elongation of torons and chiral bobbers, correspondingly.
Both 
torons and bobbers are localized in three dimensions, contain singularities, and according to the theoretical analysis within the standard phenomenological models can exist as metastable states in saturated and modulated phases of noncentrosymmetric ferromagnets. 
Their elongation into the defect-free skyrmion filament is facilitated by small anisotropic contributions making skyrmion cores negative with respect to the surrounding parental state. 
We show that isolated magnetic torons pose the same problem of compatibility with a surrounding phase as the torons in confinement-frustrated chiral nematics [I. Smalyukh \textit{et al.}, Nature Mater \textbf{9}, 139-145 (2010)].
We underline the distinct features of magnetic and liquid-crystals torons and calculate phase diagrams indicating their stability regions.
}
\end{abstract}

         
\maketitle

\section{Introduction}

\textbf{1.} \textit{Introduction.} In magnetic compounds lacking inversion symmetry, the underlying crystal structure induces a specific asymmetric exchange coupling, the so-called Dzyaloshinskii-Moriya interaction (DMI) \cite{Dz64}. 
Within a continuum approximation for magnetic properties, the DMI is expressed by Lifshitz invariants (LI) involving first derivatives of the
magnetization  $\textbf{m}$ with respect to the spatial coordinates 
\begin{equation}
\mathcal{L}^{(k)}_{i,j} = m_i \partial m_j/\partial x_k - m_j  \partial m_i/\partial x_k
\label{Lifshitz}
\end{equation}
and in a general case of cubic helimagnets has the following form \cite{Dz64,Bogdanov89}:
\begin{align}
w_D= \mathcal{L}^{(x)}_{y,z} + \mathcal{L}^{(y)}_{x,z} + \mathcal{L}^{(z)}_{x,y} = \mathbf{m}\cdot \mathrm{rot} \mathbf{m}.
\label{DMI}
\end{align}
%
%
A variety of non-collinear magnetic states (e.g., one-dimensional helicoid and conical phases) is stabilized owing to this relativistic DMI.

LIs (\ref{Lifshitz})  also help to overcome the constraints of the Hobart-Derrick theorem \cite{solitons}, and yield countable particle-like topological excitations - chiral skyrmions \cite{Bogdanov94,Bogdanov89,review,Nagaosa13}. 
Recently, skyrmion lattice states (SkL) and isolated skyrmions (ISs) were discovered in bulk crystals of chiral magnets near the magnetic
ordering temperatures \cite{Muehlbauer09,Wilhelm11,Kezsmarki15} and in nanostructures with confined geometries over larger temperature
regions\cite{Yu10,Yu11,Du15,Liang15}.


%

%

The small size and easy manipulation of skyrmions by electric fields and currents \cite{Schulz12,Jonietz10,Hsu17} generated enormous interest in their applications in information storage and processing \cite{Sampaio13,Tomasello14}.
Futhermore, complex three-dimensional internal structure of ISs and character of skyrmion-skyrmion interaction 
are imposed  by a surrounding "parental" state, e.g.,  a state  homogeneously magnetized along the field (repulsive inter-skyrmion potential)\cite{LeonovNJP16}, a conical phase with the wave vector along the field (attraction) \cite{LeonovJPCM16,LeonovAPL16} or a tilted ferromagnetic state in magnets with polar crystal structure and easy-plane anisotropy (anisotropic potential) \cite{Leonov17}, what extends even further the skyrmion functionalities in prototype spintronic devices \cite{LeonovAPL16}.

\begin{figure*}
\includegraphics[width=1.95\columnwidth]{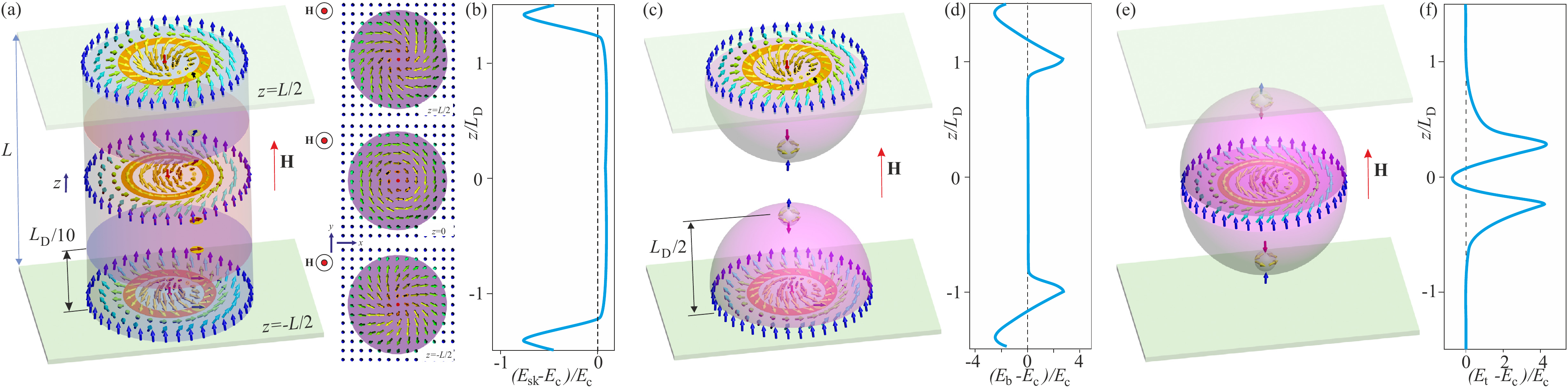}
\caption{ 
(color online) Localized particle-like states in cubic  thin-film helimagnets. (a) Skyrmionic defect-free filament  surrounded by the field-polarized or the conical phases. In thin layers of cubic helimagnets, the structure of a skyrmionic filament becomes additionally modulated in the near-surface region of width $L_D/10$. Snapshots in the transversal $xy$ plane show the structure of skyrmions in the middle layer ($z=0$, helicity value is  $\pi/2$) and for two opposite surfaces with the in-ward and the out-ward change of the azimuthal angle (helicity) of the magnetization.  (c) Schematic representation of chiral bobbers - particle-like states localized near layer surfaces and culminating in two BPs. (e) Schematic representation of  magnetic torons -  spatially localized three-dimensional skyrmions comprised by a skyrmion filament of finite length cupped with two BPs terminating its prolongation. (b), (d), (f) The energy densities of skyrmions $E_{sk}$, bobbers $E_{b}$, and torons $E_t$ averaged over the $xy$ plane and computed with respect to the energy $E_c$ of the conical phase for $h=0.5$ in the model (\ref{density}). 
\label{art} 
}
\end{figure*}

The twisting magnetization $\mathbf{m}$ in the skyrmions also matches boundary conditions at the confining surfaces of magnetic nanostructures. 
In particular in nanolayers   of cubic helimagnets \cite{PRL16,Rybakov1} with the thickness $L$ and  free boundary conditions at the lower ($z=-L/2$) and upper  surfaces ($z=L/2$, Fig. \ref{art} (a)),   the structure of skyrmions  is altered by additional chiral twists \cite{PRL16,Rybakov1}. 
The skyrmion solutions at the surfaces  are the result of the interplay between Lifshitz invariants $\mathcal{L}^{(x,y)}_{i,j}$  requiring skyrmion helicity $\gamma=\pi/2$   and $\mathcal{L}^{(z)}_{i,j}$  leading to the in-ward and out-ward rotational sense of the magnetization (see the structure of a Bloch-type skyrmion for $z=0$ and $z=\pm L/2$ in Fig. \ref{art} (a)).
Thus, the skyrmion in thin layers of cubic helimagnets could be visualized as a composite object.
The central part in the middle of the layer has higher (positive) magnetic energy as compared with the conical phase (Fig. \ref{art} (b)), see also Fig. 6 (a) in Ref. \onlinecite{Damian16}). However, the negative energy density in a narrow surface region associated with the additional twists may enable the lower total energy of skyrmions and thus lead to their thermodynamical stability.

%
Topological point defects (Bloch points \cite{Bloch}) may disrupt the smooth magnetization rotation and extend even further the variety of particle-like states in thin layers of cubic helimagnets. 
%
%
In particular due to the specific energetics exhibiting an excessive positive energy over the film thickness, isolated skyrmions  may break and transform into a pair of chiral bobbers attached to the upper and the lower surfaces of the layer (Fig. \ref{art} (c)) \cite{Rybakov2}.
Then, the structure of bobbers is balanced by the negative energy contribution stemming from additional surface twists and the positive energy due to the point defect (Fig. \ref{art} (d)). 
The Bloch point (BP) that terminates the structure of a bobber is situated at a finite distance  from the surface (Fig. \ref{art} (c)) \cite{Rybakov2}.
The chiral bobbers may provide an alternative approach for data encoding and thus be used alongside with skyrmions in magnetic solid-state memory devices \cite{Zheng2017}.

%
Interactions described by LIs (\ref{Lifshitz}) arise in other noncentrosymmetric condensed matter systems (such as antiferromagnets, chiral liquid crystals, ferroelectrics, and multiferroics) and are responsible for the formation of multidimensional solitonic states and spatially modulated phases also in these materials. 
In chiral liquid crystals (LC), a surprisingly large diversity of naturally occurring and laser-generated topologically nontrivial solitons with differently knotted nematic fields has been recently investigated \cite{Smalyukh10,Ackerman17,Ackerman16}.
In particular, a LC toron represents a localized particle  consisting of two BPs at finite distance and a convex-shaped skyrmion stretching between them (Fig. \ref{art} (e)). 
Due to the gradually varying skyrmion helicity, 
the energy density becomes negative in the toron's cross-section what is balanced by the positive energy contributions from two BPs (Fig. \ref{art} (f)). 
Thus, such a particle utilizes energetically favorable additional twist and simultaneously satisfies the boundary conditions at the confining substrates with strong surface anchoring \cite{Ackerman17b,Ackerman15} (see Supplemental Material on the details of the toron's internal structure). 
Recently, a low-voltage-driven motion of such topological LC defects with the precise control of both the direction and speed  was realized in nematic fluids \cite{Ackerman17b} what can be considered as a LC counterpart 
of a race-track memory suggested for magnetic skyrmions. 

In the present paper, we pose a problem of skyrmion nucleation in thin layers of cubic helimagnets, occuring via elongation of torons (homogeneous nucleation) and chiral bobbers (heterogeneous nucleation), since these entities are claimed to have lower activation energy as compared with IS.
Since the potential barrier that must be overcome for a particle to appear is a function of the interfacial energy with respect to the surrounding conical or homogeneous state, the heterogeneous nucleation is more common than the homogeneous one. 
In particular in Ref. \onlinecite{Rybakov2}, the spontaneous nucleation of magnetic bobbers has been observed during the simulated temperature annealing with no appearing magnetic torons.  
The LC torons, however, are easily laser generated as described in Ref. \onlinecite{Ackerman17}: (i) the realignment of the LC director $\mathbf{n}$ was locally achieved by its coupling to the optical-frequency electric field of the laser beam; (ii) alternatively, the  chiral nematic LC was locally heated to the isotropic phase of the material by a focused laser beam, so that the spontaneous appearance of torons could then be prompted upon quenching it back to the LC phase. 

We demonstrate that torons and bobbers are regular solutions of the equations describing the equilibrium states of a noncentrosymmetric system and can exist as metastable states in the saturated and modulated phases.
We argue, however, that in the isotropic case (\ref{density}) both types of skyrmion nucleation are not feasible. 
To facilitate  elongation of torons and bobbers and their subsequent transformation into the ordinary IS that pierce the layer, we apply a uniaxial anisotropy as a primary candidate making the skyrmion core negative with respect to a surrounding state.
We also construct the phase diagram of solutions indicating the stability limits of torons and elucidate their physical nature.


\textbf{2.} \textit{Model.} The standard model for magnetic states in cubic non-centrosymmetric ferromagnets is based on the energy density functional \cite{Dz64,Bak80}
\begin{equation}
w =A\,(\mathbf{grad}\,\mathbf{m})^2 + D\,\mathbf{m}\cdot \mathrm{rot}\,\mathbf{m} -\mu_0 \,M  \mathbf{m} \cdot \mathbf{H},
\label{density}
\end{equation}
including the principal interactions essential to stabilize modulated states:  the exchange stiffness with constant $A$,  Dzyaloshinskii-Moriya  coupling energy with constant $D$, and the Zeeman energy; $\mathbf{m}= (\sin\theta\cos\psi;\sin\theta\sin\psi;\cos\theta)$  is the unity vector along the magnetization vector  $\mathbf{M} = \mathbf{m} M$, and $\mathbf{H}$ is the applied magnetic field along $z-$ axis.
The film is infinite in $x-$  and $y-$ directions, i.e., we exclude any influence of the lateral sample boundaries  on the nucleation process, since inhomogeneities near sample edges readily provide nucleation centers for skyrmions (see, e.g., Fig. 3 in Ref. \onlinecite{Iwasaki} or Fig. 3 in Ref. \onlinecite{Keesman2015} showing half skyrmions at the lateral edges, which can be considered as two-dimensional defect-free counterparts of chiral bobbers \cite{Rybakov2}). On details about the discrete model used to address the chiral bobbers throughout the paper, refer to the Supplemental Material. 

%
The solutions for particle-like states in the film (Fig. \ref{art}) are derived by the Euler equations for energy functional (\ref{density}) together with the Maxwell equations and with corresponding boundary conditions.
%
%
The solutions depend on the two control parameters of the model (\ref{density}), the \textit{confinement ratio}, $\nu = L/L_D$ and the reduced value of the applied magnetic field, $h = H/H_D$ where $L_D = 4\pi A/|D|$ is the \textit{helix period} and $\mu_0 H_D = D^2/(2A M)$ is the \textit{saturation field} \cite{Bak80,Bogdanov94}. 

\begin{figure}
\includegraphics[width=0.98\columnwidth]{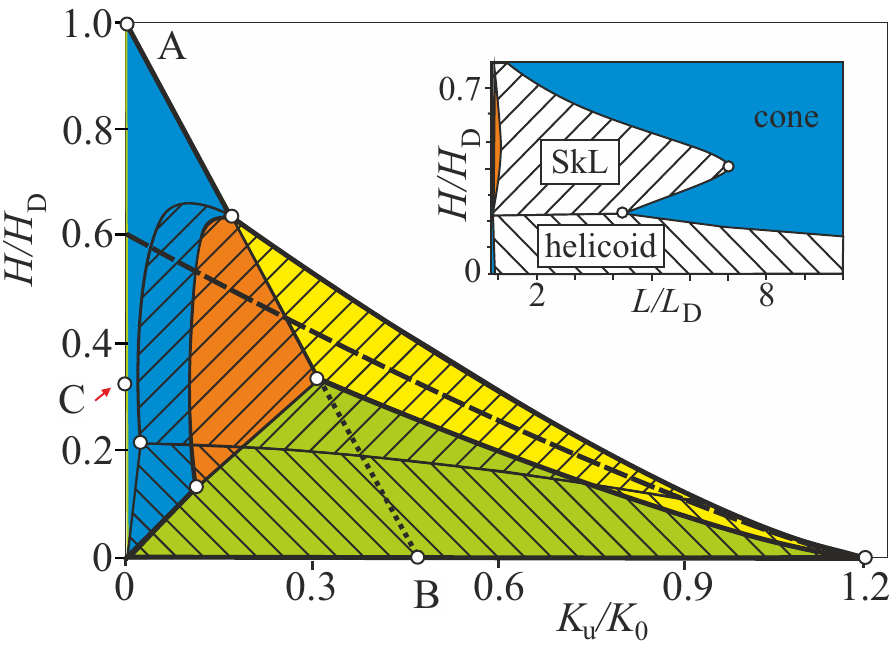}
\caption{
(color online) The diagram in coordinates $H/H_D$ - $K_u/K_0$ reflecting the internal properties of magnetic torons (see text for details). Inset shows the phase diagram for model (\ref{density}) with $K_u=0$. 
%
\label{PD}
}
\end{figure}

\textbf{3.} \textit{The phase diagrams and homogeneous nucleation.} The phase diagram of states constructed in Refs. \onlinecite{PRL16,Rybakov2016} for a thin layer within model (\ref{density}), shows vast areas of SkL and spirals stabilized due to the additional surface twists and separated by the lines of the first-order phase transition from the conical phase (see inset of Fig. \ref{PD}).
ISs within the conical phase, however, are metastable particles for all values of the confinement ratio \cite{Rybakov2} (except the small orange-shaded region for $\nu<1$ in which the energy of an IS becomes negative).
The reason lies in the specific transient region  between an IS and the conical phase (dubbed "shell" in Ref. \onlinecite{LeonovJPCM16}) that bears the positive energy density and increases linearly with the thickness. Moreover, the additional surface twist (and hence an associated negative energy) is essentially reduced in IS as compared with SkL \cite{Rybakov2}.
The energy of chiral bobbers, on the contrary, is only field-dependent and does not depend on the layer thickness \cite{Rybakov2} what makes bobbers the lowest-energy metastable states and precludes  the process of heterogeneous skyrmion nucleation. 
%
%
%
%
The homogeneous nucleation of skyrmions is also prevented within the model (\ref{density}) since a part of skyrmion with the positive energy density must be  implanted into the torons's structure, which necessarily increases its energy.
In particular in Ref. \onlinecite{Varanytsia2017} it was shown that the LC torons exist in some range of the confinement ratio. If the confinement ratio is too small, the anchoring force necessitates the toron transformation into the aligned state. On the contrary, if one tries to elongate the LC torons by increasing the confinement ratio, the torons undergo an elliptical instability towards the more stable fingerprint texture.

In the following, we supply the model (\ref{density}) with an uniaxial anisotropy of the easy-axis type \cite{Butenko10} with the easy axis $\mathbf{a}$ co-aligned with the field $\mathbf{H}$, $w_{an}= - K_u (\mathbf{m}\cdot\mathbf{a})^2$.
$K_u>0$  since for $K_u<0$ the conical phase is the global minimum in the whole region of the phase diagram \cite{Wilson14,Rowland16}.
As a solution with the period $L_D$, the conical phase exists below the critical field $H_C = H_D \left( 1- K_u/K_0 \right)$ where $K_0 = D^2 /(4A) $.
The equilibrium parameters for this cone phase  are expressed in the analytical form \cite{Bak80}  as:
\begin{eqnarray}
\theta_c = \arccos \left(H/H_C \right), \, 
 \psi_c =   2\pi z/L_D, \, \, 
\label{cone}
\end{eqnarray}
Above the critical field $H_C$,  the cone phase transforms into the saturated state with $\theta = 0$ (straight line $A-B$ in the diagram, Fig. \ref{PD}).

The diagram in Fig. \ref{PD} exhibits the following regions for  magnetic torons.
%
%
In the orange-shaded region, the energy of the core section  becomes negative with respect to the surrounding conical phase. 
Then, the magnetic toron undergoes an elongation: the longer is the distance between two Bloch points, the larger amount of the  negative energy with respect to the conical phase is "accumulated" in the core section (see Supplemental Material for details). 
Moreover, the Bloch points might be expelled altogether at the sample surfaces, thus making the skyrmion even more energetically favorable.
In the blue-shaded region, the energy density of the core section in the magnetic torons is positive with respect to the conical phase which disables their elongation. 
In the yellow-shaded region,  the magnetic torons have the negative energy in their cores, but with respect to the surrounding homogeneous state (see Supplemental Material for the details on their internal structure): thus the same process of skyrmion elongation might take place. 
The hatched regions display the thermodynamically stable hexagonal SkL and helicoids and owing to the first-order phase transition between different modulated phases  do not coincide with the colored regions for isolated torons.
Hence in a green-shaded region of the phase diagram, torons undergo an elliptical instability with respect to the thermodynamically stable helicoid. 

\begin{figure}
\includegraphics[width=0.98\columnwidth]{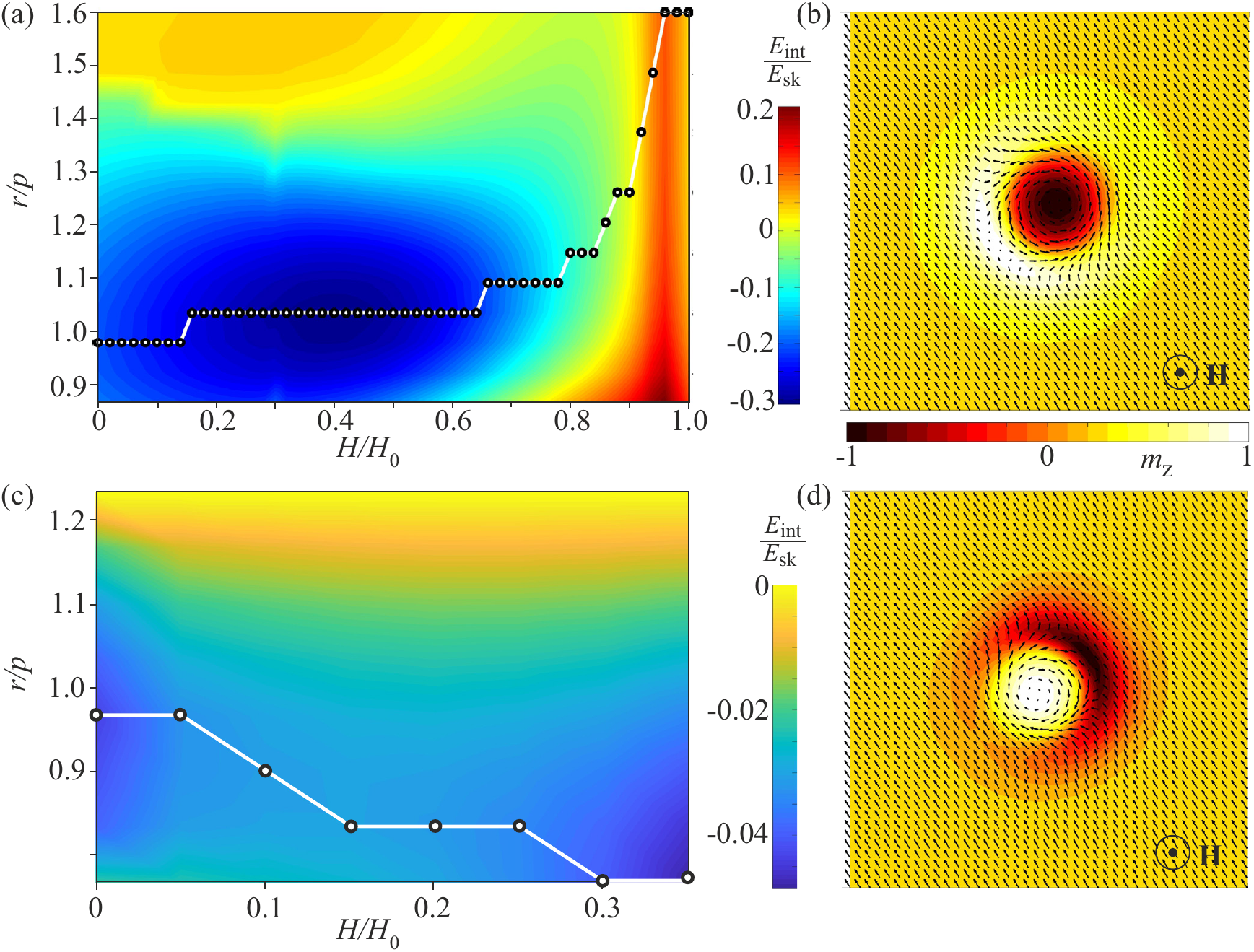}
\caption{
(color online) Reduced energy for the interaction between two non-axisymmetric skyrmion filaments (a), (c) surrounded by the conical phase, $E_{int}/E_{sk}$,  plotted as a color plot in coordinates of $r$ (the distance between the skyrmion centers) and the applied magnetic field $H/H_0$ ($K_u=0$). The attraction between skyrmions of negative (b) and positive (d) polarities has been considered.
\label{attraction}
}
\end{figure}

\begin{figure}
\includegraphics[width=0.98\columnwidth]{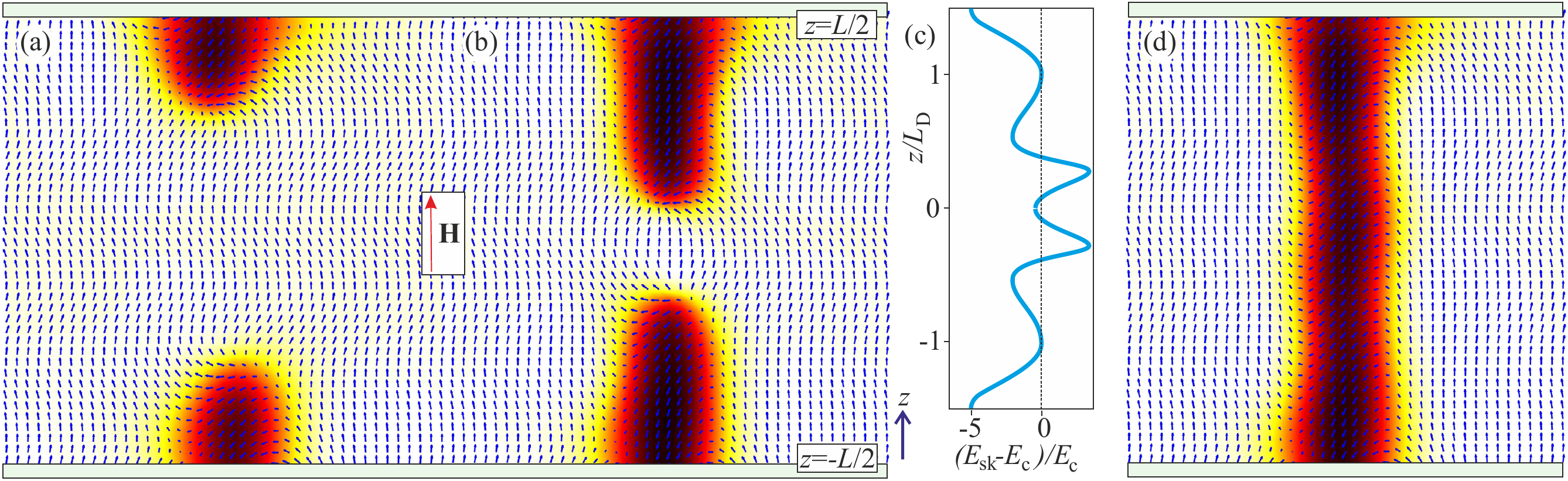}
\caption{
(color online) Heterogeneous nucleation of defect-free skyrmions (c) from chiral bobbers with defects (a) occuring via an intermediate state (b). An energy density of such an intermediate state (c) averaged over the $xy$ plane and computed with respect to the energy $E_c$ of the conical phase for $H/H_D=0.5,\,K_u/K_0=0.25$ in model (\ref{density}).
\label{bobbers}
}
\end{figure}

%
%
%
%

A process of homogeneous skyrmion nucleation from magnetic torons must be also inherent for bulk cubic helimagnets.
We treat the magnetic torons as nuclei of the first-order phase transition from the conical phase into the skyrmion lattice
which may become paramount within A-phases of bulk cubic helimagnets near the ordering temperature (e.g., in B20 magnets MnSi \cite{Muehlbauer09} and FeGe \cite{Wilhelm11}). This transition may occur via formation of magnetic torons of finite length accompanied by their elongation (e.g., due to the cubic or exchange anisotropy, which also may lead to the negative energy density of skyrmion cores) and mutual attraction. 
We stress that a softened version of the order parameter (magnetization) allows to replace the notion of localized defects as BPs by smooth but more complex geometrical adaptation of ordering with regions of suppressed order-parameter intensity \cite{nature}. 
In particular, such a method was applied to construct a new modulated phase in bulk cubic helimagnets - a squire lattice of half-skyrmions \cite{nature} -  that does not exist with the fixed value of the order parameter. 


\textbf{4.} \textit{Skyrmion-skyrmion attraction.}
In Fig.  \ref{attraction}, we plot  the interaction energy between two non-axisymmetric skyrmions  
 $E_{int}/E_{sk}$ (in units of the total equilibrium energy of an isolated asymmetric skyrmion $E_{sk}$), as a function of the distance between the skyrmion centers $r$ calculated for different values of the applied field ($K_u=0$). 
It is seen that the largest interaction energy is achieved at $H/H_D=0.4$ and equals $\approx 0.3 E_{sk}$.
In Ref. \onlinecite{Wilson14} (see Fig. 10), it was shown that in the field $H/H_D=0.4$ the difference between the energy densities of the hexagonal SkL and the cone phase is minimal. 
Thus, it was suggested that the SkL could be stabilized with respect to the cones by additional anisotropic energy contributions exactly around this field value.
With our new insight, we may add that the energy of the shell in ISs is the largest for this field value.
Therefore, while condensing into the lattice and thus by eliminating the shell, the skyrmions acquire the largest profit in their energy density. 

Note that the conical phase accommodates two types of ISs  with the magnetization in their cores either along or opposite to the field. 
At zero field, two states with the opposite polarity share the same energy (note the opposite location of the crescent-shaped region in two types of ISs with respect to their circular cores). 
In an applied magnetic field however, the skyrmions with the positive polarity may  exist only in a narrow field interval (for $K_u=0$ it is a range $0-C$ in Fig. \ref{PD}).

\textbf{5.} \textit{Heterogeneous nucleation.} 
Transition from the chiral bobbers (Fig. \ref{bobbers} (a)) to IS (Fig. \ref{bobbers} (d)) occurs via an intermediate state (Fig. \ref{bobbers} (b)) with two Bloch points located at the fixed distance from each other. 
In this case, the energy distribution in the region between two BPs (Fig. \ref{bobbers} (c)) looks qualitatively the same as for a toron (Fig. \ref{art} (f)) and is stipulated by the additional twist of the magnetization which necessarily accompanies a chiral Bloch point.
Thus, some potential barrier must be overcome to annihilate a pair of BPs which becomes also inherent for bulk helimagnets filled with torons of finite length. 
Hence we anticipate small jumps of the magnetization associated with this process in A-phases of bulk cubic helimagnets  and reminiscent the magnetic Barkhausen effect. 


To conclude, we have derived regular solutions for torons in the saturated and cone phases of cubic helimagnets.
Alongside with the chiral bobbers introduced in Ref. \onlinecite{Rybakov2}, magnetic torons may serve as nuclei of skyrmion matter: the energy of a magnetic toron may become negative in some region of the constructed phase diagrams (Fig. \ref{PD}) thus instigating its elongation.
Subsequently, defect-free ISs (or torons and bobbers of finite length) due to the mutual lateral attraction form clusters and eventually an ideal SkL. Such a process follows the definition of a nucleation-type phase transition  introduced by De Gennes\cite{deGennes} for (continuous) transitions into incommensurate modulated phases. 
A comparison of magnetic torons with their LC counterparts will facilitate their experimental investigation in thin-layer and bulk chiral helimagnets.

\section{Acknoledgements}

The authors are grateful to Ivan Smalyukh and Alex Bogdanov for useful discussions. This work was funded by JSPS Core-to-Core Program,
Advanced Research Networks (Japan) and JSPS Grant-in-Aid for Research Activity Start-up 17H06889. AOL thanks Ulrike Nitzsche for
technical assistance.

\section{Homogeneous and heterogeneous  nucleation of skyrmions in thin layers of cubic helimagnets -  Supplemental Material}

\subsection{The discrete model.}

Since a Bloch point results in an infinite exchange-energy contribution, we resort to a discrete model with magnetic moments localized on atoms, i.e., we enable not only point defects, but also collapse of skyrmions which is impossible within continuous micromagnetic models.
Thus, to investigate the solutions  for magnetic torons, we  use  the discretized version of Eq. (3):
\begin {align}
&w =  J\,\sum_{<i,j>} (\mathbf{S}_i \cdot \mathbf{S}_j ) -\sum_{i} \mathbf{H} \cdot \mathbf{S}_i - K_u S_z^2
 \nonumber\\
&- D \, \sum_{i} (\mathbf{S}_i \times \mathbf{S}_{i+\hat{x}} \cdot \hat{x}
 + \mathbf{S}_i \times \mathbf{S}_{i+\hat{y}} \cdot \hat{y} +\mathbf{S}_i \times \mathbf{S}_{i+\hat{z}} \cdot \hat{z})
\label{discrete}
\end{align}
The classical spins of the unit length are placed in the knots of a three-dimensional cubic lattice.
$<i,j>$ denotes pairs of nearest-neighbor spins.
The first term describes the ferromagnetic nearest-neighbor exchange with $J<0$ (in the numerical simulation $J=-1$ is used). 
The Dzyaloshinskii-Moriya constant $D$ defines the period of modulated structures $p$ via the following relation:  $D/J= \tan (2\pi/p)$ (thus, we use the discretized version of the DMI (2) in our forthcoming simulations).
Or vice versa, one chooses the period of the modulations $p$ (a discrete analogue of $L_D$) for the computing procedures and defines the corresponding value of the DMI constant. 
In what follows, the Dzyaloshinskii-Moriya constant is set to $0.48$ which  corresponds to one-dimensional modulations with a period of 14 lattice spacings in a zero field (for further details on the methods see Refs. [20,36]). 
%
%
The size of our numerical grid is set to $100\times 100\times L$ which is enough to accommodate an IS within the conical phase and to take into account all the subtleties of its internal structure. 
%

\subsection{The properties and 3D topology of magnetic torons}

\begin{figure*}
\includegraphics[width=1.8\columnwidth]{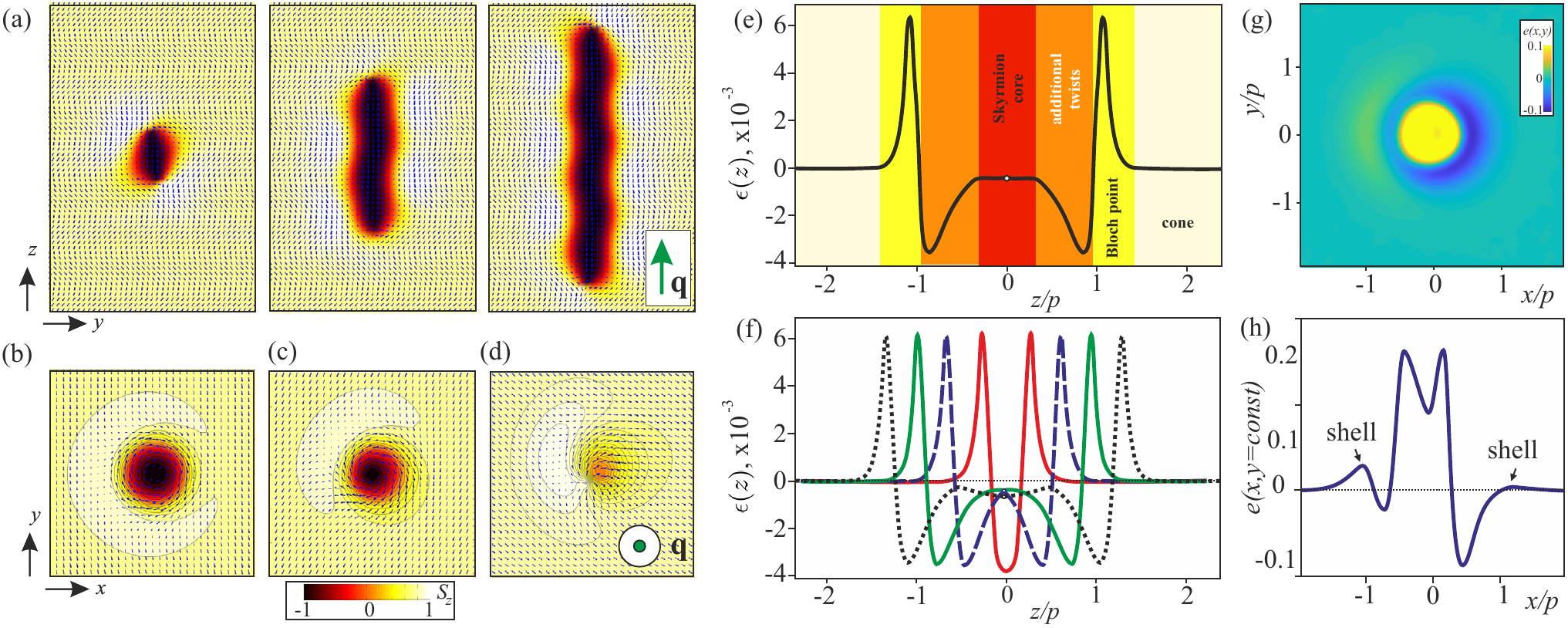}
\caption{ 
(color online). (a) Magnetic structure of isolated magnetic torons having different length along $z$ axis and surrounded by the conical phase . The color plots indicate $z$-component of the magnetization, blue arrows are projections of the magnetization on to the $yz$ plane. The green arrow shows the \textbf{q}-vector of the conical phase. (b) - (d) Composite parts of the magnetic torons shown in the transversal $xy$ plane: (b) the core section; (c) the section with the additional twists along $z$; (d) the Bloch point (see text for details). The color indicates $z$-component of the magnetization, whereas the blue arrows are the projections of the magnetization on to the $xy$-plane. (e) the energy density $\epsilon(z)$ averaged over the $xy$ plane and computed with respect to the conical phase shows the negative energy density in the region of additional twists (orange shading) and the core region (red shading) balanced by the positive energy associated with BPs (yellow region). (f) In magnetic torons with longer extension along $z$, the negative energy density $\epsilon(z)$ is gained owing to the core region. (g) The color plot for the energy density $e(x,y)$ obtained from (\ref{energyZ})  and computed with respect to the energy of the conical phase. A magnetic toron with  its core section length $p/2$ was chosen for simplicity. Such an asymmetric energy distribution implies anisotropic attracting skyrmion-skyrmion interaction. The horizontal linescan (h) of $e(x,y)$ compares the magnitudes of the shell from the left and the right sides of the core section. 
\label{fig:crank1}
}
\end{figure*}

The complex three-dimensional structure of magnetic torons which are the solutions of Eq. (\ref{discrete}) with different lengths along $z$ is stipulated by the tendency to build into the conical phase (Fig. \ref{fig:crank1} (a)). 
By this process, the magnetic torons develop a lateral transitional region towards the cones (so called "shell" [20,21]) 
and terminate their structure along $z$ by two BPs.
Three different composite sections along $z$ axis can be singled out in a magnetic toron: 

1. \textit{The  core section} is a central part of a magnetic toron. Fig. \ref{fig:crank1} (b) shows the structure of this part in the $xy$ plane: the central circular region nearly preserves the axial symmetry, whereas the transient region 
with the asymmetric crescent-like shape is formed with respect to the embedding conical state. 
This asymmetric profile of the cross-section forms a screw-like modulation along $z$ axis trying to match the conical phase at each coordinate $z$. 

The inherent properties of such non-axisymmetric skyrmion solutions with the infinite length have been extensively studied in Refs. [20,21].
It was shown that the shell has the positive energy with respect to the  cone phase and thus underlies an attractive inter-skyrmion potential. 

In the present case of magnetic torons,  the core section extends along $z$ over a finite length.
This implies that the shell, which is obtained by the energy averaging along $z$, has different magnitudes depending on the azimuthal angle. 
In Fig. \ref{fig:crank1} (g), we show the energy density (3) after integration with respect to $z$ coordinate, 
\begin{equation}
e(x,y)=(1/l)\int_l wdz.
\label{energyZ}
\end{equation} 
For this integration procedure we used a magnetic toron with the length $l=p/2$ of its core section. 
%
Such an energy distribution  underlies an anisotropic but still attracting skyrmion-skyrmion interaction.
Fig. \ref{fig:crank1} (h) shows a horizontal linescan across $e(x,y)$: it has a bump with positive energy density from the  left which is essentially lowered at  the right side.

Fig. \ref{fig:crank1} (e) shows the energy density $\epsilon(z)$ averaged over the $xy$ plane in dependence on the $z$ coordinate.
The core section of a magnetic toron is marked by the red shading. 
%

%
2. The section of magnetic torons with \textit{the additional twists}.  Fig. \ref{fig:crank1} (c) shows the structure of this part in the $xy$ plane. 
The magnetization in this section does not retain its helicity equaling $\pi/2$ as in the core section, but rather 
 undergoes an additional in-plane rotation while propagating along $z$.
The sense of the magnetization rotation is opposite while moving towards upper and lower MPs, i.e. the magnetization undergoes an out-ward (with the helicity decreasing towards $0$) and/or in-ward rotation (with the helicity increasing towards $\pi$ [24,25]). 
Due to the rotational DMI terms (2) along $z$, the additional negative energy (with the value larger than in the core section) can be "earned". 
In Fig. \ref{fig:crank1} (e) this region has orange shading. The color plot in Fig. \ref{fig:crank1} (c) displays the spin structure for the fixed value of $z$-coordinate corresponding to the minimal energy density in this orange shaded region. 
%
%

%

%
3. \textit{The Bloch points} represent singular points at which the smooth rotation of the magnetization is disrupted [26].
The structure of the Bloch points is shown in Fig. \ref{fig:crank1} (d) in the $xy$ plane.
The corresponding section of the magnetic torons is marked by the yellow shading in Fig. \ref{fig:crank1} (e).
The $z$-coordinate of the color plot corresponds to the maximal energy value in Fig. \ref{fig:crank1} (e).
%

Thus, the magnetic toron can be visualized as a composite object formed by a finite-length section of the non-axisymmetric skyrmions within the conical phase 
with the "attached" chiral bobbers at its ends. 

\subsection{toron's longitudinal stability}

\begin{figure}
\includegraphics[width=0.98\columnwidth]{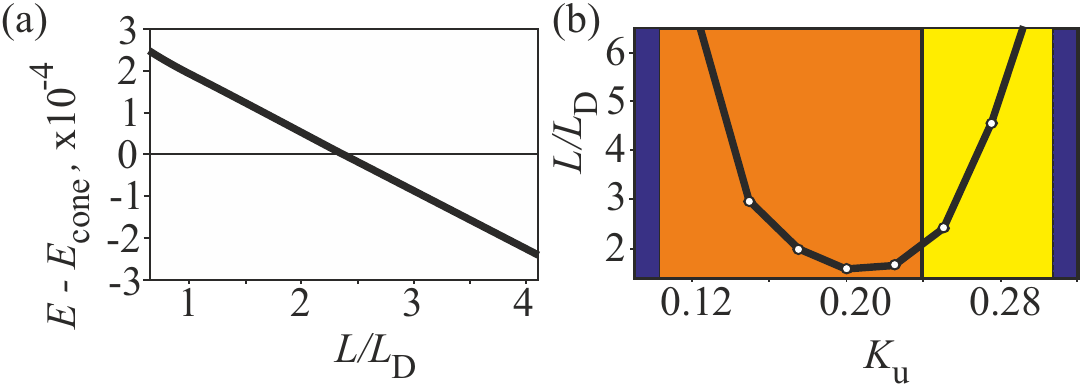}
\caption{
(color online)  (a) The total energy of a magnetic toron $E_t$ for the fixed field and the anisotropy values, $H/H_0=0.5,\,K_u/K_0=0.25$, plotted in dependence on the toron's length along $z$.  The total energy becomes negative for some critical length $l_{cr}$. (b) The critical anisotropy-dependent length of magnetic torons $l_{cr}$ for a fixed field value $H/H_0=0.5$ according to the phase diagram in Fig. 2 of the main text. The orange and yellow shading was added to distinguish between torons within the conical and the saturated states,  correspondingly.   
\label{fig:crank3}
}
\end{figure}

Fig. \ref{fig:crank1} (f) highlights the tendency of magnetic torons to elongate: it shows the averaged energy density along $z$ for magnetic torons with different longitudinal extensions. 
The critical length   of a magnetic  toron $l_{cr}$, which ensures the total negative energy (i.e. sum energy of all three composite sections) with respect to the cone, varies throughout the orange shaded region of the diagram (Fig. 2  of the main text).
The method to define the critical length $l_{cr}$ is as follows: we define the energy $E_t$ of magnetic torons with different lengths $l$ for fixed parameters $K_u$ and $H$ (Fig. \ref{fig:crank3} (b)). 
We count the length $l$ of a magnetic toron as a distance between two BPs (a distance between points with the maximal energy density in  Fig. \ref{fig:crank1} (e)).
Then, $E_t=0$ with respect to the energy of the conical phase specifies the critical length $l_{cr}$. 
%
%
Fig. \ref{fig:crank3} (b) shows such a critical length $l_{cr}$ in dependence on $K_u$ for a fixed value of the field $H/H_0=0.5$. 
The critical length $l_{cr}(K_u)$ of a magnetic toron reaches the minimal value in the middle of the orange shaded region and equals $1.65p$ for $K_u/K_0=0.2$. 
At the boundaries of the orange-shaded region, 
the critical length  diverges to infinity: such torons by loosing their BPs and consequently the section with the additional twists transform into isolated non-axisymmetric skyrmions studied in Refs. [20,21] and depicted in Fig. 1 (a).
%

%

%
The internal structure of magnetic torons in the yellow region of the phase diagram (Fig. 2  of the main text) is depicted in Fig. \ref{fig:crank5} and basically  reproduces the structure of torons within the conical phase (Fig. \ref{fig:crank1}), i.e. three composite sections may be introduced, as well (Fig. \ref{fig:crank5} (c) - (e)). 
The crucial difference, however, lies in the absence of the shell, since such magnetic torons laterally match the homogeneous background. 
Therefore, the skyrmions rather repulse each other (see for details Ref. [19]). 
At the boundary of the yellow-shaded region (Fig. 2) the energy of an isolated skyrmion (that according to Fig. \ref{fig:crank3} (b) has an infinite length) becomes zero with respect to the homogeneous state: below this line,    the skyrmions may condense into thermodynamically stable skyrmion lattice (hatched region) whereas above, they exist as metastable excitations. 
Thus, this upper boundary  is the line of the second-order phase transition between the skyrmion lattice and the homogeneous state [4,19]. 
Fig. \ref{fig:crank5} (b) shows the topological charge $Q$ along $z$ axis: 
\begin{equation}
Q=\frac{1}{4\pi}\int\int d^2r \mathbf{m}\frac{\partial \mathbf{m}}{\partial x} \frac{\partial \mathbf{m}}{\partial y}
\label{topo}
\end{equation}
It equals 1 in the core regions and the region of additional twists (red and orange shading) and decreases to 0 in the section with BPs (yellow shading).

The stable SkL with the hexagonal arrangement of skyrmions exists in the hatched region of the diagram in Fig. \ref{fig:crank3} (a).
The line of the first-order phase transition between the skyrmion lattice and the conical phase does not coincide with the line  at which the energy of an isolated magnetic toron becomes negative. 
The reason lies in the skyrmion shell with the positive energy. 
For the same reason, the first-order phase transition, that occurs between cones and SkL and involves formation of transient regions with the positive energy between corresponding phase domains, will be lagged till the linear energy density of the transient region will  be balanced by the negative surface energy density of a new phase with respect to the old one. 
%
%
%
In Ref. [S1]  in particular, it was observed that the skyrmion clusters within the conical phase have the tendency to merge into one bigger cluster:  they diminish the energy of the domain boundary with the conical phase by decreasing its linear energy density.

\subsection{Torons in chiral nematics}

\subsection{The free Frank energy}

Within the continuum theory the equilibrium distributions of the director  $\mathbf{n} (\mathbf{r})$ in confined liquid crystals are derived by solving the Euler equations for the Frank free energy density functional [S2,S3]
\begin{align}
& f (\mathbf{n})  =\frac{\mathit{K}_1}{2}(\rm{div}\,\mathbf{n})^2
+\frac{\mathit{K}_2}{2}(\bf{n}\cdot\rm{rot}\,\mathbf{n}-q_0)^2
\nonumber\\
& +\frac{\mathit{K}_3}{2}(\mathbf{n}\times\rm{rot}\,\mathbf{n})^2
-\frac{\varepsilon_a}{2}(\mathbf{n}\cdot\mathbf{E})^2 
-\frac{ \chi_a}{2}(\mathbf{n}\cdot\mathbf{H})^2.
\label{Frank}
\end{align}
Here, $K_i\, (i=1,2,3)$ and $q_0$ are elastic constants; $\mathbf{E}$ and $\mathbf{H}$ are the vectors
of applied electric and magnetic  fields, and $\varepsilon_a$ and $\chi_a$ are values of dielectric and diamagnetic anisotropies.
In the following for the sake of simplicity we  will consider only effects imposed by the magnetic field and  restrict our analysis by the one constant approximation ($K_1 = K_2 = K_3 = K$).
In this case the energy (\ref{Frank}) is reduced to the following expression
\begin{equation}
f_{\mathrm{v}}=\frac{K}{2} (\mathrm{grad}\,\mathbf{n})^2
+Kq_0\mathbf{n}\cdot{\mathrm{rot}}\,\mathbf{n}
-\frac{\chi_a}{2}(\mathbf{n}\cdot\mathbf{H})^2 .
\label{Frank2}
\end{equation}
We use here equation $ (\mathrm{grad}\,\mathbf{n})^2 = (\rm{div}\,\mathbf{n})^2 +(\bf{n}\cdot\rm{rot}\,\mathbf{n})^2 +(\mathbf{n}\times\rm{rot}\,\mathbf{n})^2 $  $+ <\mathrm{surface}$ $\mathrm{terms}>$  holding for any unity vector $\mathbf{n}$ (for details see, e.g., Ref. [S4]).

Eq. (\ref{Frank2}) implies close relations between chiral textures in both condensed matter systems -  in chiral magnets and liquid crystals.
%
%
However, in contrast to magnetic systems still favoring smooth distributions of the order parameter, liquid crystals usually form patterns composed of various types of singularities.
Defects in liquid crystals are of various dimensionalities, not only point defects, but also line and walls, and appear due to the prevalence of orientational order over positional in the applied magnetic or electric fields [S2]. 
Control and understanding of the nature of topological defects in LC is nowadays a topic of outmost interest, as the topological defects transfer topological singularities to light and could be exploited in novel devices based on singular photonics [38].
In the defects the director $\mathbf{n}$ is said to be well defined [38,S5] and the properties of defects are  well
controlled. 
These results on observations of specific skyrmion states with defects in confined cholesteric systems  can help to investigate similar structures in chiral magnets.
Liquid crystals have several advantages over magnetic systems for the investigation of various inhomogeneous structures. The system parameters can be varied over wide limits to establish necessary conditions for a given experiment; as a
rule experiments are conducted at room temperature and are comparatively simple;
the results of investigations are easily visualized [30,S6], to a degree not usually attainable
in the investigation of magnetic systems.

\subsection{Comparison of magnetic and LC torons}

Absence of a Zeeman-like term in the elastic (Frank) free energy [S2,S3] (\ref{Frank2}) leads to the disability of LC torons to elongate as described for their magnetic counterpart in Fig. \ref{fig:crank1}.
The reason is that the core section of LC torons obtained within the model (\ref{Frank2}) has the positive energy as compared with the surrounding phase (see, e.g., Ref. [39] where it was shown that in bulk cubic helimagnets the SkL is stabilized by the simultaneous effect of the magnetic field and the easy-axis anisotropy). 
Therefore a LC toron  could be visualized as two chiral bobbers attached together with the squeezed core section. 
The helicity of the director continuously changes as going from one Bloch point to another and has the value $\pi/2$ in the central plane.
%
%
%
%
The magnetic counterpart of such LC torons exists in a white region of the phase diagram, Fig. 2. 
%
%
In Ref. [S5] it was shown, however, that  the BPs comprising such LC torons do not annihilate and are bound to each other at a certain well defined distance. 
Thus a certain potential barrier is associated with the creation and annihilation of torons.

According to the phase diagram in Fig. 2, without a Zeeman term in (3) one gets solutions for LC torons for $K_u/K_0>1.24$, i.e., a case of a strong anchoring.
For a weaker anchoring, one may enter the region of the spiral thermodynamical stability (hatched region in Fig. 2).
In this case, the defect-free localized solutions - spherulites [S7]  - may occur, such solutions however may undergo an elliptical instability towards spirals.

\begin{figure}
\includegraphics[width=0.95\columnwidth]{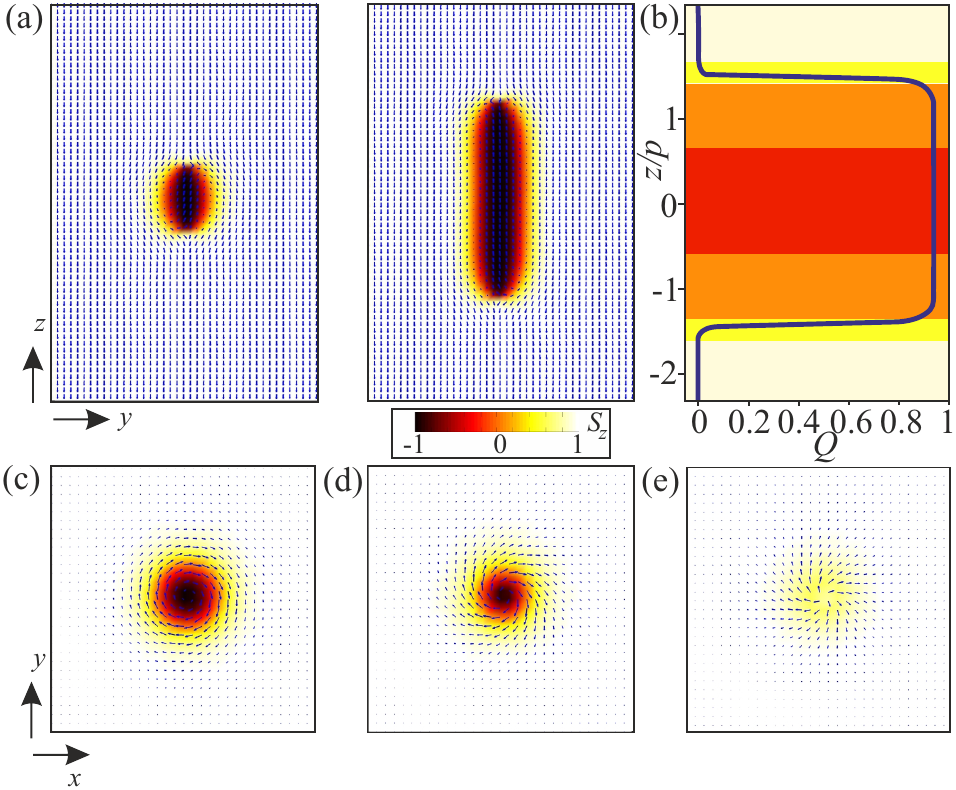}
\caption{ 
(a) Magnetic structure of isolated magnetic torons having different lengths along $z$ axis and surrounded by the saturated phase. The color plot indicates $z$-component of the magnetization, blue arrows are projections of the magnetization on to the $yz$ plane. (b) topological charge $Q$ (\ref{topo}) calculated in $xy$ planes in dependence on $z$ coordinate. The color shading corresponds to the color scheme in Fig. S1 (e). (c) - (e) Composite parts of a magnetic toron shown in the transversal $xy$ plane: (b) the core region; (c) the region with the additional twists along $z$; (d) the Bloch point (see text for details). The color indicates $z$-component of the magnetization, whereas the blue arrows are the projections of the magnetization on to the $xy$-plane.  
\label{fig:crank5}
}
\end{figure}

Interestingly, the LC torons with the non-axisymmetric core section (realized in Fig. \ref{fig:crank1} (b) in the applied magnetic field) may be realized as a result of the competition between the surface anchoring and the electric-field term in (\ref{Frank}): whereas the surface anchoring tends to orient the director $\mathbf{n} (\mathbf{r})$ perpendicular to the confining glass plates, the electric field $\mathbf{E}$ with the negative dielectric anisotropy $\varepsilon_a$ - parallel to them (see in particular Fig. 1 in Ref. [S8]). 
As  a result of such an interplay, one creates an analogue of the conical phase around a skyrmion that induces an attractive toron-toron interaction and experimentally observed toron chains [S9]. Moreover, a directional motion of such skyrmions is possible as a response to modulated electric fields: when alternating current with some frequency is applied to a confined skyrmion with the axisymmetric structure (Fig. \ref{fig:crank5} (c)), it transforms to a non-axisymmetric solution (\ref{fig:crank1} (b)) back and forth  thus inducing a squirming motion.

In general in chiral nematic LC, point and line defects spontaneously occur as a result of symmetry-breaking phase transitions and versatile 3D topological solitons might be  stabilized. 
And as an alternative to SkL, different ordered structures of defects could be organized.
In Ref. [S5], it was shown that using a scanning laser generation system, one can program a focused laser beam to generate periodic lattices formed by the metastable LC torons.
In chiral magnets, also different periodic arrangements of BPs  have been considered. In particular,  the monopole-antimonopole pairs were arranged in the form of a lattice in which they are connected by the skyrmion strings [S10].
It was shown that such a lattice has non-trivial transport properties which may result in particular in a novel magnetoresistivity effect as applied for MnGe [S10].

\textbf{Supplementary references}

[S1] J. C. Loudon, A. O. Leonov, A. N. Bogdanov, M. Ciomaga Hatnean, and G. Balakrishnan, arxiv: 1704.06876.

[S2] P. Oswald and P. Pieranski, \textit{Nematic and Cholesteric Liquid Crystals: Concepts and Physical Properties Illustrated by Experiments} (CRC Press, 2005).

[S3] M. Kleman, O. D. Lavrentovich, \textit{Soft matter physics: an introduction}, (Springer-Vertag, New York, 2003).

[S4] A. Hubert, R. Sch\"{a}fer, \textit{Magnetic Domains}  (Springer, Berlin, 1998).

[S5] P. J. Ackerman, Z. Qi, I. I. Smalyukh, Phys. Rev. E \textbf{86}, 021703 (2010).

[S6] H. Sohn, P. J. Ackerman, T. J. Boyle, G. H. Sheetah, B. Fornberg, I. Smalyukh, arxiv: 1711.01354. 

[S7] A. O. Leonov, I. E. Dragunov, U. K. R\"o\ss ler, A. N. Bogdanov, Phys. Rev. E \textbf{90}, 042502 (2014). 

[S8] P. J. Ackerman, T. Boyle, I. I. Smalyukh, Nat. Commun. \textbf{8}, 673 (2017).

[S9] P. J. Ackerman, J. van de Lagemaat, and I. I. Smalyukh, Nat. Commun. \textbf{6}, 6012 (2015).

[S10] X.-X. Zhang, A. S. Mishchenko, G. De Filippis, N. Nagaosa, Phys.Rev. B \textbf{94} 174428 (2016);  X.-X. Zhang and  N. Nagaosa, New J. Phys. \textbf{19} 043012 (2017).

\end{document}